\documentclass[letterpaper]{article} 
\usepackage{aaai24}  
\usepackage{times}  
\usepackage{helvet}  
\usepackage{courier}  
\usepackage[hyphens]{url}  
\usepackage{graphicx} 
\usepackage{natbib}  
\usepackage{caption} 
\usepackage{algorithm}
\usepackage{algorithmic}
\usepackage{amsmath}
\usepackage{multirow}
\usepackage{bm}
\usepackage{makecell}
\usepackage{svg}
\usepackage{amssymb}
\usepackage{multirow}
\usepackage{booktabs}
\usepackage{graphicx}
\usepackage{subfig} 
\usepackage{float}
\usepackage{color}
\usepackage{colortbl}

\def \etal {et al.}

\definecolor{mygray}{gray}{.9}

\usepackage{newfloat}
\usepackage{listings}
\DeclareCaptionStyle{ruled}{labelfont=normalfont,labelsep=colon,strut=off} 
\floatstyle{ruled}
\newfloat{listing}{tb}{lst}{}
\floatname{listing}{Listing}
\title{A User-Friendly Framework for Generating Model-Preferred Prompts in Text-to-Image Synthesis}
\author{
    Nailei Hei\textsuperscript{\rm 1},    
    Qianyu Guo\textsuperscript{\rm 3},
    Zihao Wang\textsuperscript{\rm 4},
    Yan Wang\textsuperscript{\rm 1}\equalcontrib,
    Haofen Wang\textsuperscript{\rm 5}\equalcontrib,
    Wenqiang Zhang\textsuperscript{\rm 1,\rm 2,\rm 3}\equalcontrib
}
\affiliations{
    \textsuperscript{\rm 1}Shanghai Engineering Research Center of AI \& Robotics, Academy for Engineering \& Technology, Fudan University \\
    \textsuperscript{\rm 2}Engineering Research Center of Robotics, Ministry of Education, Academy for Engineering \& Technology, Fudan University \\
    \textsuperscript{\rm 3}Shanghai Key Lab of Intelligent Information Processing, School of Computer Science, Fudan University\\
    \textsuperscript{\rm 4}Tongji University \\
    \textsuperscript{\rm 5}College of Design and Innovation, Tongji University \\
    nlhei22@m.fudan.edu.cn, qyguo20@fudan.edu.cn, zihaowang26@outlook.com\\
    yanwang19@fudan.edu.cn, carter.whfcarter@gmail.com, wqzhang@fudan.edu.cn
    
}

\usepackage{bibentry}

\begin{document}

\maketitle

\begin{abstract}
    Well-designed prompts have demonstrated the potential to guide text-to-image models in generating amazing images. 
    Although existing prompt engineering methods can provide high-level guidance, it is challenging for novice users to achieve the desired results by manually entering prompts due to a discrepancy between novice-user-input prompts and the model-preferred prompts. To bridge the distribution gap between user input behavior and model training datasets, we first construct a novel Coarse-Fine Granularity Prompts dataset (CFP) and propose a novel User-Friendly Fine-Grained Text Generation framework (UF-FGTG) for automated prompt optimization.
    For CFP, we construct a novel dataset for text-to-image tasks that combines coarse and fine-grained prompts to facilitate the development of automated prompt generation methods.
    For UF-FGTG, we propose a novel framework that automatically translates user-input prompts into model-preferred prompts.
    Specifically, we propose a prompt refiner that continually rewrites prompts to empower users to select results that align with their unique needs.  
    Meanwhile, we integrate image-related loss functions from the text-to-image model into the training process of text generation to generate model-preferred prompts. 
    Additionally, we propose an adaptive feature extraction module to ensure diversity in the generated results. 
    Experiments demonstrate that our approach is capable of generating more visually appealing and diverse images than previous state-of-the-art methods, achieving an average improvement of $5\%$ across six quality and aesthetic metrics.
    Data and code are available at \url{https://github.com/Naylenv/UF-FGTG}.
\end{abstract}

\section{Introduction}
\begin{figure}[t]
    \centering
    \resizebox{\columnwidth}{!}{        
        \includegraphics[scale=1]{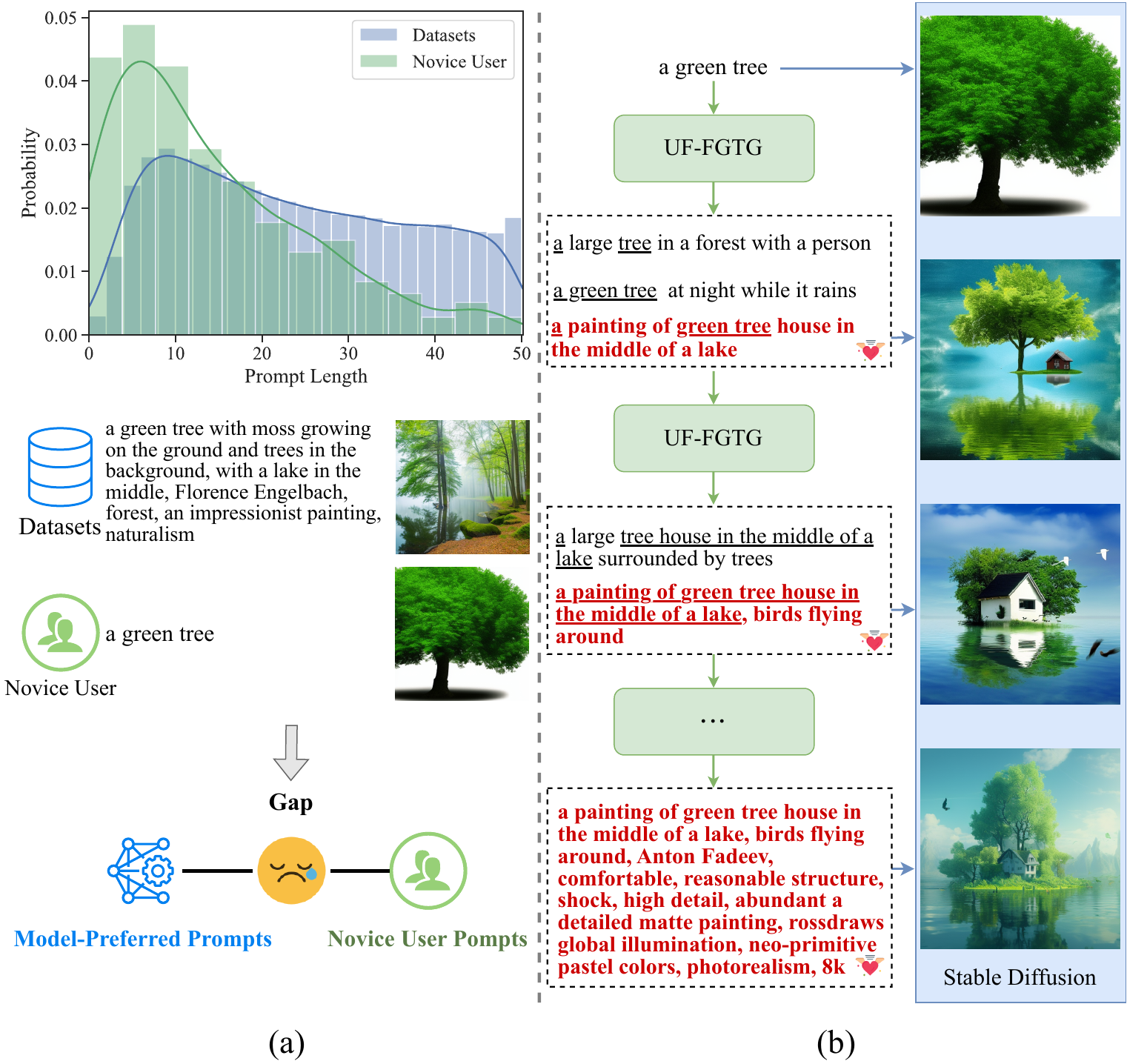}
    }
    \caption{
        (a) We uncover an inconsistency in the word length distribution between prompts in the text-to-image training dataset and those provided by novice users, leading to a misalignment between model-preferred prompts and novice user prompts.
        (b) Our proposed UF-FGTG continually rewrites prompts, allowing users to select results of interest based on their needs until satisfied.
    }
    \label{fig:introduction_sample}
\end{figure}
Generative foundation models, including language models and text-to-image models, can be prompted to follow user instructions.
Recent advancements in text-to-image synthesis such as Stable Diffusion (SD)~\cite{SD_2022_CVPR} and Midjourney~\cite{midjourney} have facilitated the generation of high-fidelity images based on text prompts.
Concurrently, recent studies have revealed that prompt design plays a crucial role in determining the quality of the generated images~\cite{Large-scale_2023_ICIUI,Opal_2022_SUIST}.
Adjusting the prompt to better reflect the user's intentions can lead to superior results.
This issue is particularly pronounced in text-to-image models, as the capacity of their text encoders is relatively limited, such as CLIP text encoder~\cite{CLIP_2021_ICML} in Stable Diffusion.
Empirical observations have also shown that common user inputs are often insufficient to produce aesthetically pleasing images using current models~\cite{google_prompt_2022_arxiv}.
\par
However, previous research has primarily focused on manually designing prompts for specific text-to-image models~\cite{prompt-engineering_2022_CHI,best_prompts_2023_sigir}, typically adding some modifiers to the original prompts.
While these studies have provided valuable insights, they are labor-intensive and only offer high-level suggestions, failing to offer personalized recommendations for users seeking specific aesthetics. Novice users, who lack experience in prompt writing and familiarity with relevant keywords, face significant challenges in achieving their desired results due to this.
Therefore, it is essential to develop a method that can automatically rewrite prompts, thereby assisting novice users in generating model-preferred prompts.
\par
As shown in Fig.~\ref{fig:introduction_sample}(a), we analyze the probability distribution of prompt word lengths in text-to-image training datasets as compared to those actually used by novice users. Specifically, we employ DiffusionDB~\cite{datasets:diffusiondb}, a large-scale dataset frequently employed for training in text-to-image tasks, as our analysis dataset. Following~\cite{HPS_2023_ICCV}, we use DiscordChatExporter~\cite{discord}
to collect novice user prompts from the \textit{`dreambot'} channel on the Stable Diffusion Discord. Our analysis reveals a tendency among novice users to input short, coarse-grained prompts, contrasting with the long, fine-grained prompts used in model training. We believe that this gap results in a discrepancy between the intentions of novice users and the prompts that the model prefers.
\par
Traditional methods for converting user-input prompts into model-preferred prompts rely on generative language models. However, existing generative language models such as GPT~\cite{gpt-2} and T5~\cite{T5_ML_2020} are restricted to uni-modal text information during training, which constrains their ability to generate genuinely model-preferred prompts. To overcome this limitation and generate high-quality images in text-to-image tasks, the need for a multi-modal training framework is clearly highlighted.
\par
To solve the above issues, we propose the \textbf{C}oarse-\textbf{F}ine Granularity \textbf{P}rompts dataset (CFP), a collection of 81,910 data instances from popular text-to-image community. Specifically, we refer to prompts in Lexica.art~\cite{datasets:santana_gustavostastable-diffusion-prompts_2022} as fine-grained prompts. Then we generate corresponding images from the fine-grained prompts and use a summarization model~\cite{summarization} to produce coarse-grained prompts, thereby creating a triplet dataset.
\par
Building on our CFP dataset, we propose a novel \textbf{U}ser-\textbf{F}riendly \textbf{F}ine-\textbf{G}rained \textbf{T}ext \textbf{G}eneration framework (UF-FGTG) for automated prompt
optimization. 
Specifically, we first propose a prompt refiner, which transforms coarse-grained prompts into fine-grained prompts. 
Secondly, we incorporate image-related loss functions in text-to-image tasks, ensuring the generated fine-grained prompts as model-preferred prompts. 
Thirdly, nearly every word in a text-to-image task prompt can find corresponding semantics in the generated image. However, many stylistic details of an image, particularly those described in short texts, are not represented. To prevent the generation in a fixed style, we propose an adaptive feature extraction module to ensure the diversity of the generated results.
As shown in Fig.~\ref{fig:introduction_sample}(b), our UF-FGTG continually refines prompts, enabling users to select outcomes of interest as per their requirements until satisfaction is achieved. Through extensive experiments on the CFP dataset, we demonstrate the effectiveness of our proposed method on both quantitative and qualitative measures.
\par
Our major contributions are as follows:
\begin{itemize}
    \item We propose a novel Coarse-Fine Granularity Prompts dataset (CFP), a unique triplet dataset designed to bridge the gap between user behavior and model-preferred prompts. To the best of our knowledge,  CFP is the first dataset that comprises fine-grained prompts with corresponding images, as well as coarse-grained prompts.
    \item We propose a novel training framework for text generation in text-to-image tasks, which transforms coarse-grained prompts into a fine-grained prompt feature space, named User-Friendly Fine-Grained Text Generation (UF-FGTG).
    \item  We propose an adaptive feature extraction module that aligns prompt features with adaptive image features to prevent the generation of monotonous style results, ensuring diversity in the generated results.
\end{itemize}
\section{Related Work}
\begin{figure*}[t]
    \centering
    \resizebox{\textwidth}{!}{
        \includegraphics[scale=1]{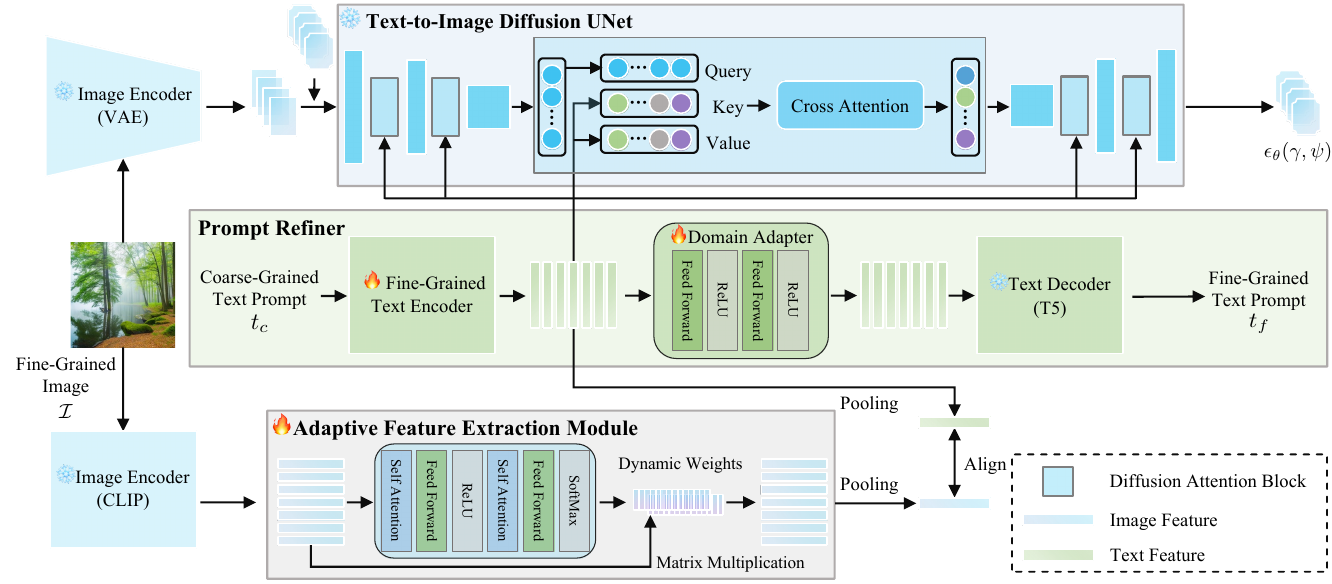}
    }
    \caption{The architecture of our User-Friendly Fine-Grained Text Generation (UF-FGTG) Framework. The crux of the text generation network is the prompt refiner, which primarily comprises a fine-grained text encoder $E_T$ and a text decoder $D_E$. The encoder $E_T$ transforms coarse-grained prompt features into fine-grained prompt features. This process is supervised by fine-grained text $T_F$, a Stable Diffusion model $\epsilon_{\theta}$, and an adaptive feature extraction module $\mathbf{\mathcal{N}}$, ensuring the generated fine-grained prompts are not only model-preferred but also diverse. During inference process, only the prompt refiner is necessary.}
    \label{fig:model}
\end{figure*}
\subsection{Text-to-Image Generative Models}
Text-to-image generative models enable users to create images based on textual input. Researchers have explored a variety of architectures to enhance image quality, including autoregressive models~\cite{Zero-shot-text-to-image_2021_PMLP}, generative adversarial networks (GANs)~\cite{StyleGAN-T_2023_ICLR}, and diffusion models~\cite{SD_2022_CVPR}.
Several subsequent works, including 
DALL-E-2~\cite{DELL-E-2_2022_arxiv}, 
GLIDE~\cite{GLIDE_2022_PMLP}, 
Imagen~\cite{Imagen_2022_NIPS}, 
and Stable Diffusion~\cite{SD_2022_CVPR}, have brought the magic of text-to-image generation to public attention. 
Among these models, Stable Diffusion stands out as an open-source model with an active user community. Although text-to-image models can generate impressive images, they demand high-quality input prompts. Novice users, lacking experience in prompt writing and unfamiliar with relevant keywords, may still struggle to generate the desired images.
\subsection{Prompt Engineering}
In the specific field of text-to-image generative models, prompt engineering research is nascent, which aims to use carefully selected and combined sentences to achieve a specific visual style in synthesized images~\cite{taxonomy_prompt_modifiers_2022_arxiv}.
The input texts, known as the prompts, direct the model to generate specific images.
Manual prompt engineering is a natural method for optimizing prompts.
Pavlichenko \etal~\shortcite{best_prompts_2023_sigir};
Oppenlaender \etal~\shortcite{taxonomy_prompt_modifiers_2022_arxiv};
Liu \etal~\shortcite{prompt-engineering_2022_CHI}
explore how to find model-preferred prompts manually.
While manual prompt engineering can lead to significant progress, the process of designing prompts requires time and experience, and may not always yield optimal results.
Therefore, various methods have focused on automatically searching for prompts through 
mining~\cite{mining_2020_TACL}, 
parsing~\cite{parsing_2021_arxiv}, 
text generation~\cite{google_prompt_2022_arxiv} 
and LLMs~\cite{Collaborative_Generative_AI_2023_EMNLP,I_spy_2023_ACL}.
Additionally, many previous works have focused on gradient-based prompt learning methods, such as ~\cite{CoOp_2022_IJCV, hard_prompts_2023_arxiv}. However, these gradient-based methods are not human-readable, which may hinder their application in human-AI collaboration.
\par
Our work is closely related to prior research in the field of prompt engineering. For example, Google's recent study~\cite{google_prompt_2022_arxiv} introduced a reinforcement learning-based approach to prompt training.  However, their strategy is essentially a training methodology that can be applied to other models. Another related work by Wang \etal~\shortcite{reprompt_2023_CHI} is limited by its reliance on manual design and may not be easily applicable to other domains.
In this work, we propose a user-friendly prompt engineering framework that is data-driven and interpretable for text-to-image generation.
\section{Coarse-Fine Granularity Prompts Dataset}
\paragraph{Motivation.}
Existing datasets primarily depend on fine-grained prompts and corresponding images for model training~\cite{datasets:diffusiondb}. However, in real-world scenarios, users frequently input coarse-grained prompts, leading to a disparity between the model's training and inference phases. Addressing this discrepancy to align native-user-input prompts with model-preferred prompts is crucial. To bridge this gap, we construct the Coarse-Fine Granularity Prompts dataset (CFP).
\paragraph{Fine-Grained Prompts Collection.}
We build a coarse-fine granularity prompts dataset based on Lexica.art~\cite{datasets:santana_gustavostastable-diffusion-prompts_2022}, which consists of 81,910 fine-grained prompts filtered and extracted from user communities.
\paragraph{Coarse-Grained Prompts and Fine-Grained Images.}
For each fine-grained prompt obtained, following~\cite{datasets:diffusiondb}, we use Stable Diffusion-v2.1~\cite{SD_2022_CVPR} to generate a corresponding image. The parameters used for image generation include \textit{``step", ``seed", ``height", ``width", ``CFG scale", and ``sampler"}.
Additionally, we employ BART~\cite{BART_2022_ACL} as a summarization model~\cite{summarization} to generate coarse-grained prompts of three different lengths: 1-5 tokens, 6-10 tokens, and 11-15 tokens. During training, one of these coarse-grained prompts is selected randomly.
\paragraph{Data Format.}
Finally, we obtain a total of 81,910 data instances, each consisting of one fine-grained prompt, one fine-grained image, and three coarse-grained prompts. We split 73,718 data pairs as the training set and 8,192 data instances as the testing set.
\paragraph{NSFW Contents.}
Following~\cite{HPS_2023_ICCV} and~\cite{laion_nips_2022}, we observe that a small subset of data instances may contain NSFW (not safe for work) content. To avoid the potential harm caused by such content, we employ an NSFW detector~\cite{NSFW_text_classifier} to filter out fine-grained prompts that contain NSFW elements. We suggest using only the data with scores below 0.9, which amounts to a total of 79,447 data instances.
\section{UF-FGTG Framework}
\paragraph{Motivation.}
Existing text generation methods are uni-modal, which can ensure the transformation of coarse-grained prompts into fine-grained ones, but cannot guarantee model-preferred prompts. To solve this issue, we propose the User-Friendly Fine-Grained Text Generation framework (UF-FGTG), which has the capability to transform coarse-grained prompts into the feature space of fine-grained, model-preferred prompts, thereby generating high-quality images. More specifically, we propose a prompt refiner, which has the ability to transform coarse-grained prompts into fine-grained prompts. To ensure the generated prompts are model-preferred, we incorporate image-related supervision from Stable Diffusion. Additionally, we propose an adaptive feature extraction module that ensures diversity in the generated results.
\subsection{Framework Overview}
Fig.~\ref{fig:model} presents an overview of our framework, specifically designed for prompt generation. We take the Stable Diffusion~\cite{SD_2022_CVPR} as an example to introduce our methodology. Our model is trained using triplet datasets, which we denote as $\mathcal{S} = \{(t_c, t_f, \mathcal{I})\}$. Here, $t_c$ stands for coarse-grained prompts, while $t_f$ represents fine-grained prompts. The symbol $\mathcal{I}$ corresponds to the images that are associated with the fine-grained prompts.
\par
The core of our framework is prompt refiner mainly composed of fine-grained text encoder and text decoder, which are designed to transform the input coarse-grained prompts into fine-grained prompts. For the first time, we incorporate image-related supervision from the Stable Diffusion process to generate model-preferred prompts, ensuring that the generated fine-grained prompts align with the model-preferred prompts. Additionally, we observe that while nearly every word in a prompt can find its corresponding semantics in the generated image, many stylistic details, particularly in short texts, are not adequately represented. This issue confines our UF-FGTG to generate images in a specific style. To tackle this problem, we propose an adaptive feature extraction module that aligns prompt features with adaptive image features, thereby ensuring diversity in the generated results.
\par
During inference, the user inputs coarse-grained prompts and utilizes the prompt refiner within the framework to generate fine-grained prompts preferred by the model.
\subsection{Text-to-Image Diffusion Model}
\paragraph{Stable Diffusion.}
It consists of three main components: an autoencoder $\mathcal{A}$, a text time-conditional UNet denoising model $\epsilon_{\theta}$, and a CLIP fine-grained text encoder $E_{T}$. The autoencoder $\mathcal{A}$ includes a VAE encoder $\mathcal{E}$ and a VAE decoder $\mathcal{D}$, while the CLIP text encoder $E_{T}$ accepts text prompts $t$ as input.
The encoder $\mathcal{E}$ transforms an image $\mathcal{I} \in \mathbb{R}^{3 \times H \times W}$ into a lower-dimensional latent space in $\mathbb{R}^{4 \times h \times w}$, where $h=H/8$ and $w=W/8$. Conversely, the decoder $\mathcal{D}$ carries out the inverse operation, decoding a latent variable into the pixel space.
\par
Our goal is to generate fine-grained prompts $t_f$ from the coarse-grained prompts $t_c$ provided by the user while ensuring that the feature space can be understood by the UNet denoising model $\epsilon_{\theta}$, which is designed to generate fine-grained prompts that are model-preferred prompts.
To achieve this,
we refer to the $\epsilon_{\theta}$ convolutional input as the spatial input $\gamma$ (e.g., $z_t$) since convolutions preserve the spatial structure, and to the attention conditioning input as $\psi$ (e.g., $[\tau, E_{T}(T)]$).
We train the text encoder $E_{T}$ by minimizing the loss function defined as follows:
\begin{equation}
    \mathcal{L}_\mathrm{mse} = \mathbb{E}_{\mathcal{E}(\mathcal{I}), t_c, \epsilon \sim \mathcal{N}(0,1),\tau} \lVert \epsilon - \epsilon_{\theta}(\gamma,\psi) \rVert_2^2,
    \label{eq:diffusion_loss}
\end{equation}
where $\tau$ represents the diffusing time step, $\gamma = z_\tau$, $z_\tau$ is the encoded image $\mathcal{E}(\mathcal{I})$ where we stochastically add Gaussian noise $\epsilon \sim \mathcal{N}(0,1)$, and $\psi=\left[\tau;E_{T}(t_c)\right]$.

\subsection{Prompt Refiner}
It consists of three components: a fine-grained text encoder $E_{T}$, a text decoder $D_{T}$, and a domain adapter $Q$. We utilize the CLIP model as the fine-grained text encoder and the T5 model~\cite{T5_ML_2020} as the text decoder to articulate our methodology.
\paragraph{Fine-Grained Text Encoder.}
CLIP~\cite{CLIP_2021_ICML} is a vision-language model that aligns visual and textual information within a shared embedding space. CLIP consists of a visual encoder $E_{V}$ and a text encoder $E_{T}$. These encoders independently generate feature representations $E_{V}(\mathcal{I}) \in \mathbb{R}^{n}$ for an input image $\mathcal{I}$, and $E_{T}(L(t)) \in \mathbb{R}^{n}$ for the corresponding text $t$. Here, $n$ represents the dimensionality of the embedding space in CLIP, and $L$ denotes the embedding lookup layer that maps each tokenized word $t$ to its respective token embedding in space $\mathcal{W}$.
\par
In the original Stable Diffusion model, the CLIP text encoder only has the capability for text encoding. However, our fine-grained text encoder can transform the feature space from coarse-grained prompts $t_c$ to model-preferred fine-grained prompts $t_f$,
by concurrently employing the fine-grained prompt-related loss and the image-related loss supervision from the Stable Diffusion.
In subsequent sections, we employ a language model to decode these features into human-readable prompts.
\paragraph{Text Decoder.}
The objective of our fine-grained text encoder is to transform coarse-grained prompt features into model-preferred fine-grained prompt features. Additionally, we implement a feature domain adapter $Q$, which utilizes a Multilayer Perceptron (MLP) to project CLIP text features onto the T5 text features space. Simultaneously, we employ the T5 model as a text feature decoder, denoted as $D_T$, to generate the final human-readable fine-grained prompts.
\par
More specifically, the fine-grained text encoder is initialized using OpenCLIP~\cite{openclip_cvpr_2023} derived from the Stable Diffusion model~\cite{SD_2022_CVPR}. The text decoder $D_T$ is initialized with a FLAN-T5~\cite{flan-t5} pretrained generative language model. The fine-grained prompts $t_f$ are utilized as training labels. The training objective is to minimize the log-likelihood by leveraging the teacher forcing technique~\cite{SFT_loss_1989}:

\begin{equation}
    \mathcal{L}_{\mathrm{sft}}=\mathbb{E}_{(t_c, t_f)\sim \mathcal{S}} \log p(t_f \mid Q(E_T(t_c))).
\end{equation}
\subsection{Adaptive Feature Extraction Module}
\begin{figure*}[htb]
    \centering
    \resizebox{\textwidth}{!}{
        \includegraphics[scale=1]{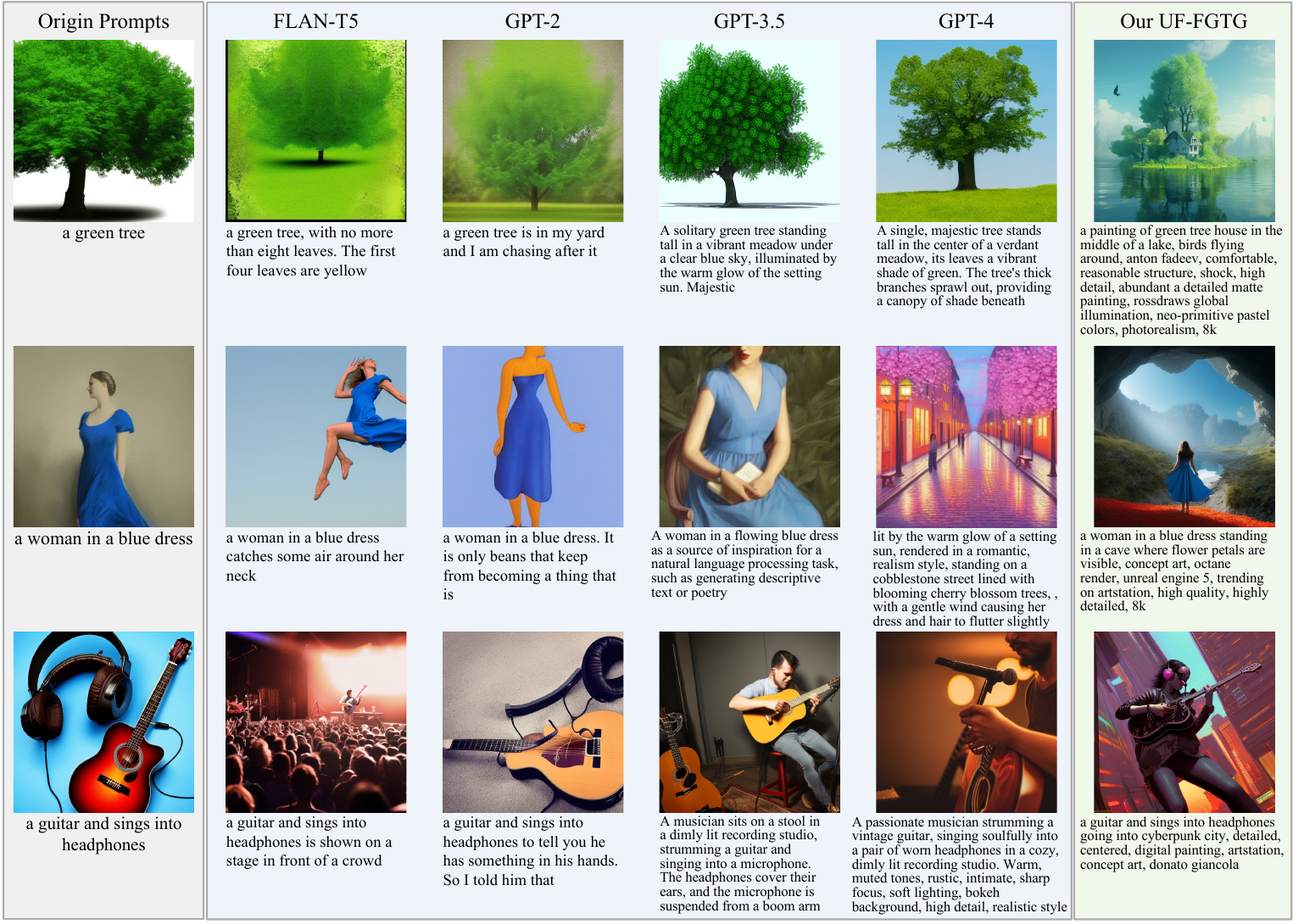}
    }
    \caption{ Comparison of prompts generated by FLAN-T5, GPT-2, GPT3.5, GPT-4, and our UF-FGTG, with corresponding images generated by Stable Diffusion-v2.1.
    }
    \label{fig:quantitative}
\end{figure*}
Through text-to-image model and prompt refiner in our framework, the fine-grained text encoder $E_{T}$ transforms coarse-grained prompt features into model-preferred fine-grained prompt features. While nearly every word in a prompt can find its corresponding semantics in the generated image, many stylistic details in the image are difficult to reflect in the prompts, especially in short ones. For instance, a coarse-grained text such as \textit{``a green tree"} might align with \textit{a sunny forest scene} image, which, through direct training, could lead to generated results adhering to a uniform style, thereby reducing diversity. To ensure diversity in the generated results, we propose an adaptive feature extraction module that adaptively extracts image features.
\par
Fig.~\ref{fig:model} illustrates the architecture of our adaptive feature extraction module. This module aims to predict the soft dynamic weights of image representations. Specifically, we take the image $\mathcal{I}$ extracted by the CLIP image encoder $E_V$ as the input to the dynamic weight network. The dynamic weight network consists of self-attention layer, feed-forward layer, and ReLU activation functions. Ultimately, the dynamic weights $w$ are obtained through a SoftMax function. These weights are then applied to weight pixel-wise features through matrix multiplication. This method enables the automatic learning of the most suitable and relevant representation from the image feature.
\par
The features $E_T(t_c)$, which are extracted by the fine-grained text encoder, and the features $\mathbf{\mathcal{N}}(E_V(\mathcal{I}))$, generated by the CLIP image encoder and adaptive feature extraction module $\mathbf{\mathcal{N}}$, have the same dimensions. As suggested by~\cite{clip_loss_2023_iccv}, we use the CLIP-Enhance loss function to evaluate the similarity between the prompt features and image features. The training objective is to minimize the CLIP-Enhance loss, where $B$ represents the batch size:
\begin{equation}
    \mathcal{L}_\mathrm{clip} = -\frac{1}{B} \log{\Sigma \frac{e^{\cos (E_T(t_c)_{i}, \mathbf{\mathcal{N}}(E_V(\mathcal{I}))_{i})}}{\Sigma_{j \neq i} e^{\cos (E_T(t_c)_{i}, \mathbf{\mathcal{N}}(E_V(\mathcal{I}))_{j})}}}.
\end{equation}
\subsection{Loss Function}
The overall loss function is a weighted sum of $\mathcal{L}_\mathrm{mse}$, $\mathcal{L}_\mathrm{sft}$ and $\mathcal{L}_\mathrm{clip}$. In our experiments, we set the trade-off hyper-parameters $\alpha_{1}$ and $\alpha_{2}$ to 0.1.
\begin{equation}
    \mathcal{L}=\mathcal{L}_\mathrm{mse}+\alpha_{1} \mathcal{L}_\mathrm{sft}+\alpha_{2} \mathcal{L}_\mathrm{clip}.
\end{equation}

\section{Experiments}
\begin{table}[htb]

    \setlength\tabcolsep{3pt}
    \centering
    \resizebox{1.00\columnwidth}{!}{

        \begin{tabular}{c|cccc|cc}
            \toprule
            Model          & \begin{tabular}[c]{@{}c@{}}NIMA\\ -TID$\uparrow$\end{tabular} & \begin{tabular}[c]{@{}c@{}}MUSIQ\\ -KonIQ$\uparrow$\end{tabular} & \begin{tabular}[c]{@{}c@{}}DB-\\ CNN$\uparrow$\end{tabular} & \begin{tabular}[c]{@{}c@{}}TReS\\ $\uparrow$\end{tabular} & \begin{tabular}[c]{@{}c@{}}NIMA\\ -AVA$\uparrow$\end{tabular} & \begin{tabular}[c]{@{}c@{}}MUSIQ\\ -AVA$\uparrow$\end{tabular} \\ \midrule
            GPT-2          & 5.37                                                          & 68.79                                                            & 61.19                                                       & 76.80                                                     & 5.17                                                          & 5.60                                                           \\
            FLAN-T5        & 5.40                                                          & 68.61                                                            & 61.18                                                       & 77.11                                                     & 5.19                                                          & 5.60                                                           \\
            GPT-3.5        & 5.61                                                          & 66.74                                                            & 60.63                                                       & 75.28                                                     & 5.30                                                          & 5.80                                                           \\
            GPT-4          & 5.54                                                          & 67.29                                                            & 60.82                                                       & 76.06                                                     & 5.22                                                          & 5.85                                                           \\
            GPT-2*         & 5.62                                                          & 69.55                                                            & 62.40                                                       & 80.81                                                     & 5.32                                                          & 5.81                                                           \\
            FLAN-T5*       & 5.59                                                          & 69.59                                                            & 63.19                                                       & 80.90                                                     & 5.29                                                          & 5.84                                                           \\ \midrule
            UF-FGTG*       & \textbf{5.73}                                                 & \textbf{69.74}                                                   & \textbf{65.21}                                              & \textbf{83.34}                                            & \textbf{5.48}                                                 & \textbf{5.97}                                                  \\ \midrule
            \rowcolor{mygray}
            \textit{CFP-C} & \textit{5.46}                                                 & \textit{68.76}                                                   & \textit{62.05}                                              & \textit{78.06}                                            & \textit{5.25}                                                 & \textit{5.66}                                                  \\
            \rowcolor{mygray}
            \textit{CFP-F} & \textit{5.81}                                                 & \textit{70.15}                                                   & \textit{66.97}                                              & \textit{84.18}                                            & \textit{5.62}                                                 & \textit{6.07}                                                  \\ \bottomrule
        \end{tabular}
    }
    \caption{Image quality \& aesthetic assessment. All methods first generate fine-grained prompts from coarse-grained prompts and then evaluate the generated images using Stable Diffusion-v2.1. “*” means trained on Coarse-Fine Granularity Prompts dataset (CFP). \textit{CFP-C} and \textit{CFP-F} denote the coarse-grained and fine-grained prompts in the CFP dataset.}
    \label{table:quality_and_aesthetic_assessment}
\end{table}
\subsection{Experimental Setting}
\paragraph{Implementation Details.}
We conduct our experiments on NVIDIA A100 GPUs. During training, we train the fine-grained text encoder, domain adapter, and adaptive feature extraction module on our CFP dataset for 100 epochs, using the AdamW optimizer~\cite{AdamW}, a learning rate of 5e-5, and a batch size of 16.
\par
In line with the Stable Diffusion-v2.1, our fine-grained text encoder is initialized with OpenCLIP~\cite{openclip_cvpr_2023}. The text decoder is initialized using FLAN-T5-base~\cite{flan-t5}, while the image encoder employs the OpenCLIP that is paired with the fine-grained text encoder. This consistent approach to initialization ensures the compatibility and effectiveness of our proposed model.
\paragraph{Generation Strategy.}
In the subsequent experiments, we use the following default configuration: during prompt generation, we generate fine-grained prompts using 6-10 tokens of coarse-grained prompts. Following~\cite{top-p-k}, we employ a strategy that combines $\text{Top-p}$ and $\text{Top-K}$, which set $p$ to 0.95 and $K$ to 50. For the image generation phase, we utilize Stable Diffusion-v2.1, setting the CFG scale to 7, and perform 50 denoising steps using the Euler Ancestral sampler~\cite{euler_a_nips_2022}.
\subsection{Qualitative Comparison}
In Fig.~\ref{fig:quantitative}, we visualize the generation results from various models, including GPT-2~\cite{gpt-2}, FLAN-T5~\cite{flan-t5}, GPT-3.5~\cite{gpt-4}, and GPT-4~\cite{gpt-4}. First, we rewrite the coarse-grained prompts, setting the max tokens to 20.
Then, we generate images using Stable Diffusion-v2.1. Our method is capable of producing visually appealing images.
Furthermore, we note that traditional language models, including GPT-2 and FLAN-T5, struggle to comprehend the format of model-preferred prompts in text-to-image tasks.
Even when we provide ChatGPT with a prompts format during generation
it still fails to produce satisfactory results~\cite{chatgpt_diffusion}. For example, in the case of \textit{``a woman in a blue dress''}, GPT-4 modifies the original semantics, leading to a significant deviation in the generated result from the original content.
In most cases, GPT-2 and FLAN-T5 can only generate short text, even when we attempt to increase the maximum token count.
In summary, these issues originate from language models' limited grasp of image information in text-to-image tasks and their unfamiliarity with the structure of preferred prompts.
Our method effectively addresses these issues.
\begin{table}[t]
    \centering
    \setlength\tabcolsep{3pt}
    \resizebox{1.00\columnwidth}{!}{

        \begin{tabular}{@{}l|cccc|cc@{}}
            
            \toprule
            Model                                          & \begin{tabular}[c]{@{}c@{}}NIMA\\ -TID$\uparrow$\end{tabular} & \begin{tabular}[c]{@{}c@{}}MUSIQ\\ -KonIQ$\uparrow$\end{tabular} & \begin{tabular}[c]{@{}c@{}}DB-\\ CNN$\uparrow$\end{tabular} & \begin{tabular}[c]{@{}c@{}}TReS\\ $\uparrow$\end{tabular} & \begin{tabular}[c]{@{}c@{}}NIMA\\ -AVA$\uparrow$\end{tabular} & \begin{tabular}[c]{@{}c@{}}MUSIQ\\ -AVA$\uparrow$\end{tabular} \\ \midrule
            \small{wo $\mathcal{L}_\mathrm{mse,clip}$} & 5.48                                                          & 67.23                                                            & 62.34                                                       & 78.21                                                     & 5.21                                                          & 5.64                                                           \\
            wo $\mathcal{L}_\mathrm{mse}$                         & 5.53                                                          & 68.65                                                            & 64.01                                                       & 80.92                                                     & 5.35                                                          & 5.73                                                           \\
            wo $\mathcal{L}_\mathrm{clip}$                        & 5.68                                                          & 69.32                                                            & 64.74                                                       & 82.24                                                     & 5.41                                                          & 5.89                                                           \\ \midrule
            UF-FGTG                                        & \textbf{5.73}                                                 & \textbf{69.74}                                                   & \textbf{65.21}                                              & \textbf{83.34}                                            & \textbf{5.48}                                                 & \textbf{5.97}                                                  \\ \bottomrule
        \end{tabular}
    }
    \caption{Impact of Stable Diffusion model ($\mathcal{L}_\mathrm{mse}$) and adaptive feature extraction module ($\mathcal{L}_\mathrm{clip}$)  in our UF-FGTG.}
    \label{table:loss_functions}
\end{table}

\begin{figure}[b]
    \centering
    \resizebox{0.75\columnwidth}{!}{
        \includegraphics{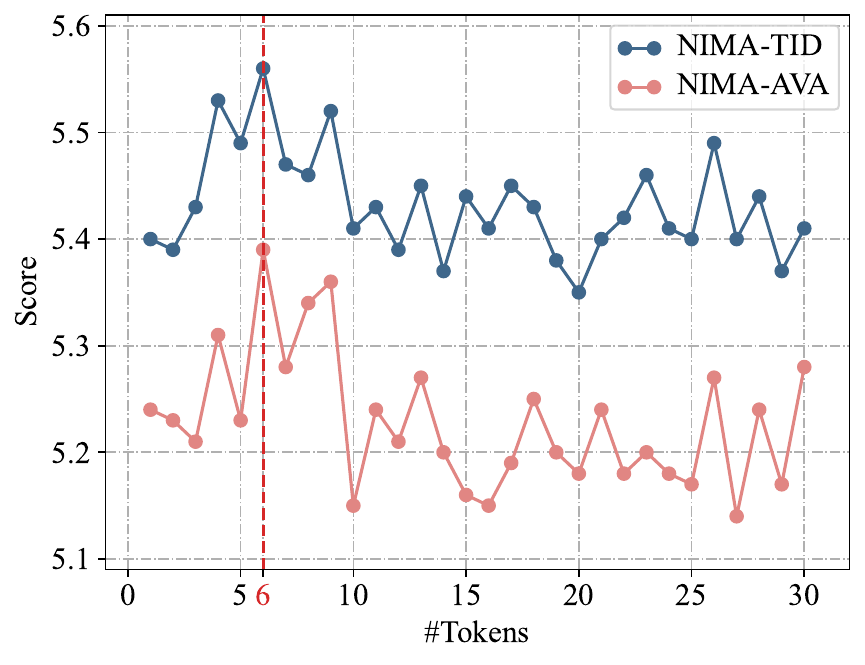}
    }
    \caption{Ablation study on prompt length, showing both NIMA-TID and NIMA-AVA score as length increases.}
    \label{fig:prompt_length}
\end{figure}

\subsection{Quantitative Comparison}
\paragraph{Evaluation Metrics.}
\begin{figure}[htb]
    \centering
    \resizebox{\columnwidth}{!}{
        \includegraphics[scale=1]{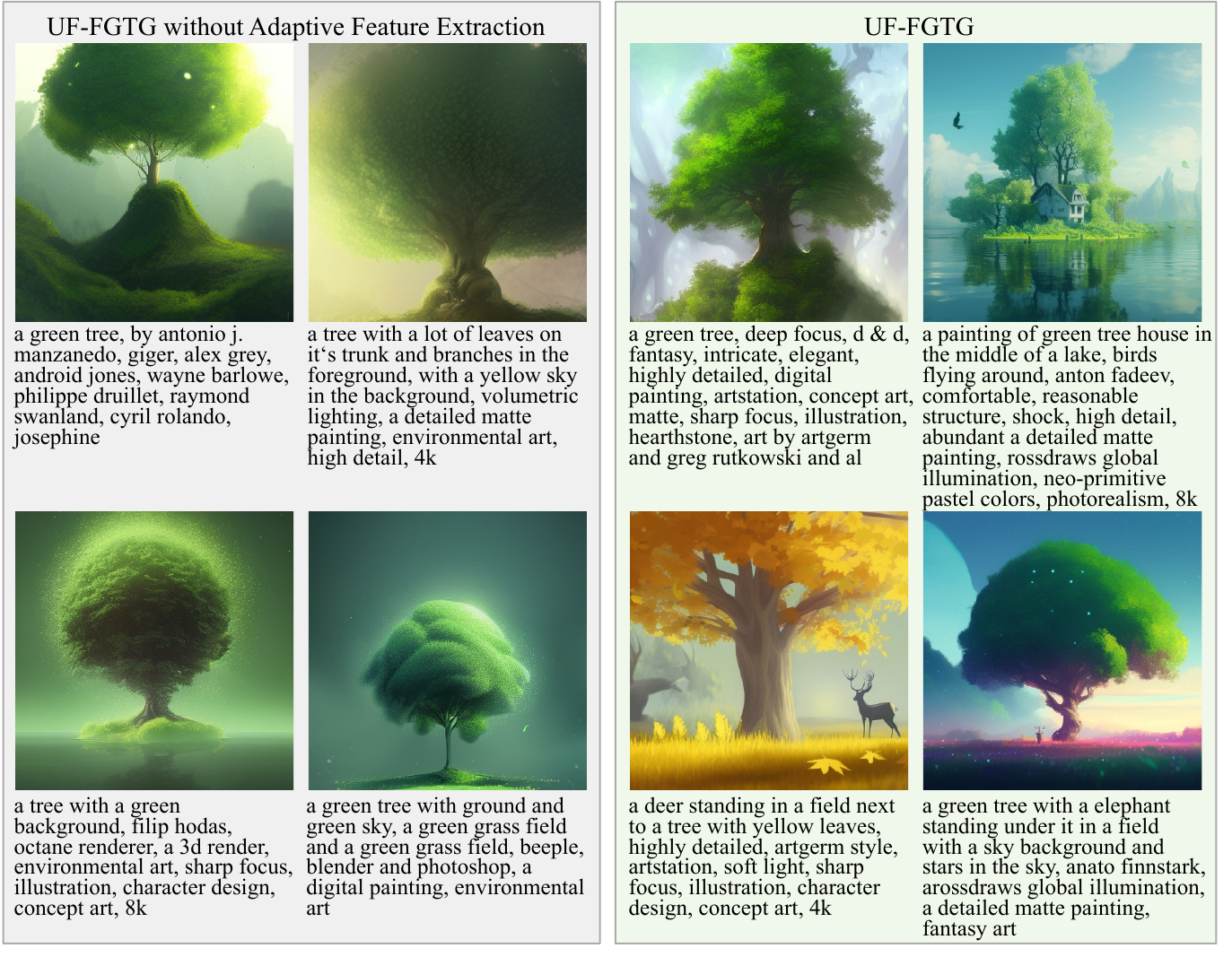}
    }
    \caption{The adaptive feature extraction module enhances the diversity of the results. Origin prompts: \textit{``a green tree''}.}
    \label{fig:effect_of_DFSM}
\end{figure}
Following~\cite{AGIQA-3K_2023} and~\cite{TISE_2022_ECCV}, we quantitatively assess the image quality and aesthetic, using the non-reference metrics.
Specifically, we choose NIMA~\cite{nima}, MUSIQ~\cite{musiq}, DB-CNN~\cite{DBCNN}, and TReS~\cite{tres}. We use NIMA-TID, MUSIQ-KonIQ, DB-CNN, and TReS for image quality assessment, while NIMA-AVA and MUSIQ-AVA are used for image aesthetic assessment. NIMA-AVA and MUSIQ-AVA are trained on AVA~\cite{ava}, MUSIQ-KonIQ and DB-CNN are trained on KonIQ-10K~\cite{koniq10k}, and NIMA-TID is trained on TID2013~\cite{tid2013}, following Py-IQA~\cite{pyiqa}.
\par

Tab.~\ref{table:quality_and_aesthetic_assessment}
showcases the performance of various generative language models in image quality and aesthetic evaluation, with our method consistently surpassing other approaches across all six metrics, achieving an average improvement of $5\%$.
This indicates that our method not only generates high-quality but substantial aesthetic images.
Two primary reasons contribute to this phenomenon:
(1) Previous uni-model text generation methods for fine-grained prompts neglect image information in text-to-image tasks, leading to subpar generative performance. Nonetheless, fine-tuning these methods with our CFP dataset yields enhanced results.
(2) We observe that the text feature space in text-to-image tasks is merely a subset of the high-dimensional space employed in text generation. Our method maps the high-dimensional feature space in text generation into a low-dimensional one suitable for text-to-image tasks. This ensures that fine-grained prompts, derived from coarse-grained prompts, are model-preferred and correspond to high-quality images.
\subsection{Ablation Study}
\paragraph{Effect of Loss Functions.}
In Tab.~\ref{table:loss_functions}, we study the model performance by varying its configuration. Our framework consists of three pipelines: text-to-image model, prompt refiner, and adaptive feature extraction module. It can be controlled by three loss functions, $\mathcal{L}_\mathrm{mse}$, $\mathcal{L}_\mathrm{sft}$, and $\mathcal{L}_\mathrm{clip}$, to determine whether to add a particular pipeline. The results indicate that $\mathcal{L}_\mathrm{sft}$ and $\mathcal{L}_\mathrm{clip}$ are indispensable for text generation in text-to-image tasks.
\paragraph{Prompt Length.}
Following~\cite{hard_prompts_2023_arxiv}, we further ablate the optimal number of tokens. In Fig.~\ref{fig:prompt_length},
we observe that when using Stable Diffusion for image generation, longer prompts do not always lead to better image quality and aesthetic assessment. This phenomenon may be attributed to overfitting caused by longer prompts. Our experience indicates that additional prompts with a length of 6 produce the most generalizable performance.
\paragraph{Effect of Adaptive Feature Extraction Module.}
\begin{figure}[htb]
    \centering
    \resizebox{\columnwidth}{!}{
        \includegraphics[scale=1]{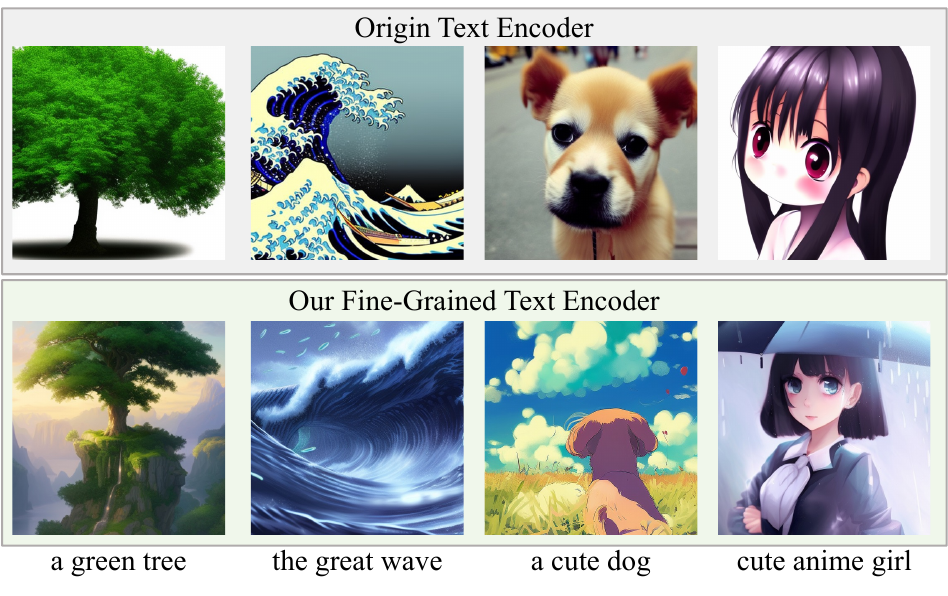}
    }
    \caption{
        Origin text encoder vs our fine-grained text encoder
        in Stable Diffusion-v2.1.
    }
    \label{fig:new_text_encoder}
\end{figure}
In Fig.~\ref{fig:effect_of_DFSM}, we demonstrate the increased diversity of generation results achieved by the adaptive feature extraction module. Without this module, the model tends to produce results with a singular style. However, by incorporating the adaptive feature extraction module, the model is capable of generating a variety of results. This diversity is attributed to our adaptive extraction of image features, which enhances the heterogeneity of the fine-grained prompt features.
\subsection{Applications}
\paragraph{Prompt Generation.}
The inference process for text generation is independent of its training phase. Our model proposes two recommended strategies for inference: (1) As shown in Fig.~\ref{fig:introduction_sample}(b), the model generates three results concurrently, with each outcome further producing six tokens based on the preceding prompts, iterating this process until user satisfaction is achieved. (2) As shown in Fig.~\ref{fig:quantitative}, the model aims to generate comprehensive prompts (setting max token to 20 or 50). Both strategies use the original Stable Diffusion model for image generation.
\paragraph{A Plug-and-Play Module in Stable Diffusion.}
Our method trains a fine-grained text encoder with the ability to map coarse-grained prompts to a fine-grained prompt feature space. As shown in Fig.~\ref{fig:new_text_encoder}, this allows it to fully supplant the encoding-only text encoder in the original Stable Diffusion model. Furthermore, we observe that when the input prompt extends to a certain length, the model tends to generate prompts such as \textit{``4k resolution", ``highly detailed" and ``best quality"}. Although these prompts are not semantically explicit, they can improve the quality of the generated images~\cite{best_prompts_2023_sigir}.
This suggests that our approach consistently projects any user-provided input prompt into a feature space aligned with fine-grained prompts, resulting in enhanced image generation quality.

\section{Conclusion}
In this paper, we propose the Coarse-Fine Granularity Prompts dataset (CFP), which enables the study of the gap between user behavior and model-preferred prompts. We also propose the User-Friendly Fine-Grained Text Generation (UF-FGTG) framework, which automatically translates user-input prompts into model-preferred prompts while incorporating image-related loss functions and an adaptive feature extraction module for improved diversity in generation results. Our experiments demonstrate that our method achieves state-of-the-art performance on both quantitative and qualitative measures, significantly advancing the field of text-to-image synthesis by providing a user-friendly method for automated prompt optimization.

\clearpage
\section{Acknowledgments}
This work was supported by
Fundamental Research Funds for the Central Universities (No.22120230032),
National Nature Science Foundation of China (No.62176185, No.62072112),
Scientific and Technological Innovation action Plan of Shanghai Science and Technology Committee (No.22511102202),
and China Postdoctoral Science Foundation (2023M730647, 2023TQ0075).
\bibliography{aaai24}
\clearpage
\section{Overview}
In this Appendix, we provide more details of CFP and UF-FGTG organized as follows:
\par
In Sec.~\textbf{Appendix A}, we provide further details on the CFP Dataset, which corresponds to Sec. \textbf{Coarse-Fine Granularity Prompts Dataset} in the main body of this paper.
\par 
In Sec.~\textbf{Appendix B}, we present additional visualizations, which correspond to Sec. \textbf{Qualitative Comparison} in the main body of this paper.
\par 
In Sec.~\textbf{Appendix C}, we describe the types of descriptions used for experimental analysis with ChatGPT, specifically GPT-3.5 and GPT-4.
\par 
In Sec.~\textbf{Appendix D}, we analyze the transferability of our model. Although our model is trained on Stable Diffusion-v2.1, the generated prompts can produce high-quality images on other text-to-image models as well.

\section{Appendix A: CFP Dataset details}
\label{sec:appendix_a}
\begin{figure}[htp]
    \centering
    \resizebox{\columnwidth}{!}{
        \includegraphics[scale=1]{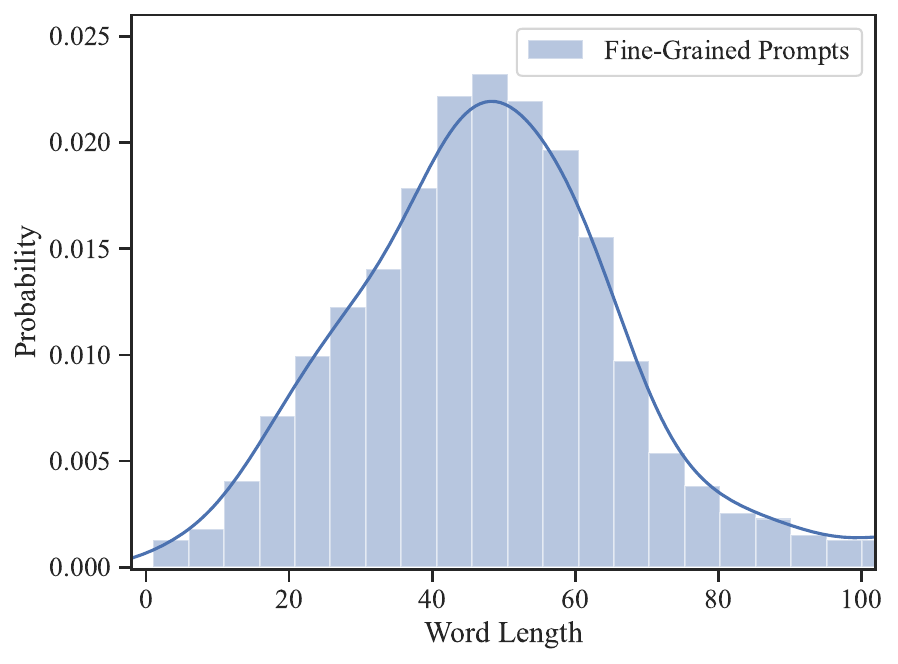}
    }
    \caption{Word length distribution in fine-grained prompts of the CFP dataset.}
    \label{fig:CFP-F}
\end{figure}

\begin{figure*}[htp]
    \centering
    \resizebox{\textwidth}{!}{
        \includegraphics[scale=1]{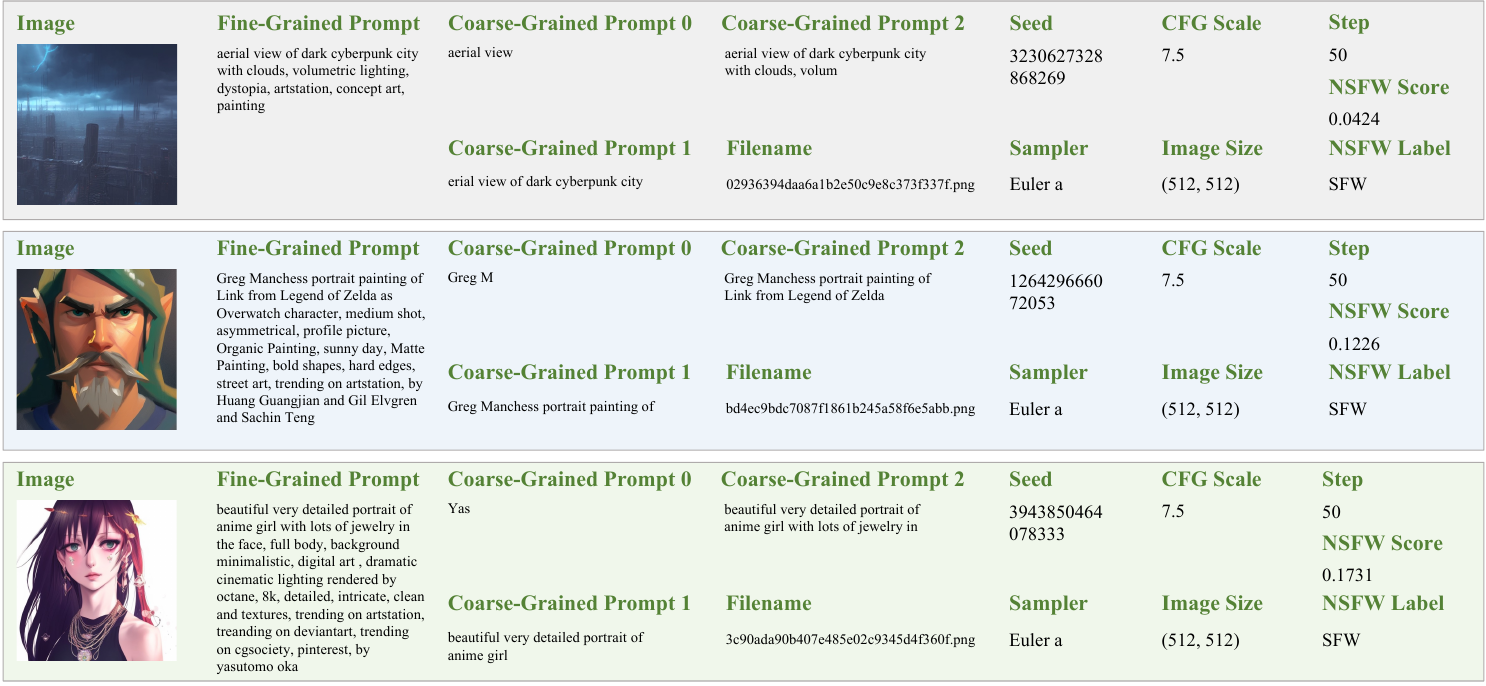}
    }
    \vspace{-1em}
    \caption{The CFP dataset comprises 81,910 data instances, each containing a fine-grained image, a fine-grained prompt, and three coarse-grained prompts of different lengths (1-5 tokens, 6-10 tokens, 11-15 tokens), as well as all model hyperparameters: \textit{`seed'}, \textit{`step'}, \textit{`CFG scale'}, \textit{`sampler'}, and \textit{`image size'}. Additionally, each image has a unique filename. To assist researchers in identifying potentially unsafe or harmful content, we employ state-of-the-art models to calculate an NSFW score for each fine-grained prompt.}
    \vspace{-1em}
    \label{fig:CFP}
\end{figure*}

We construct 
Coarse-Fine Granularity Prompts dataset (CFP) based on Lexica.art~\cite{datasets:santana_gustavostastable-diffusion-prompts_2022}, and organize the data according to DiffusionDB~\cite{datasets:diffusiondb}, as illustrated in Fig.~\ref{fig:CFP}. To ensure that the fine-grained prompts in our dataset align with the distribution of training data, we analyze the distribution of fine-grained prompt word lengths, as shown in Fig.~\ref{fig:CFP-F}.
\par 
Finally, we obtain a total of 81,910 data instances, each consisting of one fine-grained prompt, one fine-grained image, and three coarse-grained prompts. We split 73,718 data pairs as the training set and 8,192 data instances as the testing set. We also employ an NSFW detector~\cite{NSFW_text_classifier} to filter out fine-grained prompts that contain NSFW elements. We suggest using only the data with scores below 0.9, which amounts to a total of 79,447 data instances. In Table 1 of the main text, we quantitatively analyze the fine-grained prompts and find that they are superior to coarse-grained prompts in terms of both quality and aesthetics. This further demonstrates the necessity of our dataset, and the fine-grained prompts are preferred by models.
\section{Appendix B: Additional Visualizations}
\label{sec:appendix_b}
\begin{figure*}[htb]
    \centering
    \resizebox{\textwidth}{!}{
        \includegraphics[scale=1]{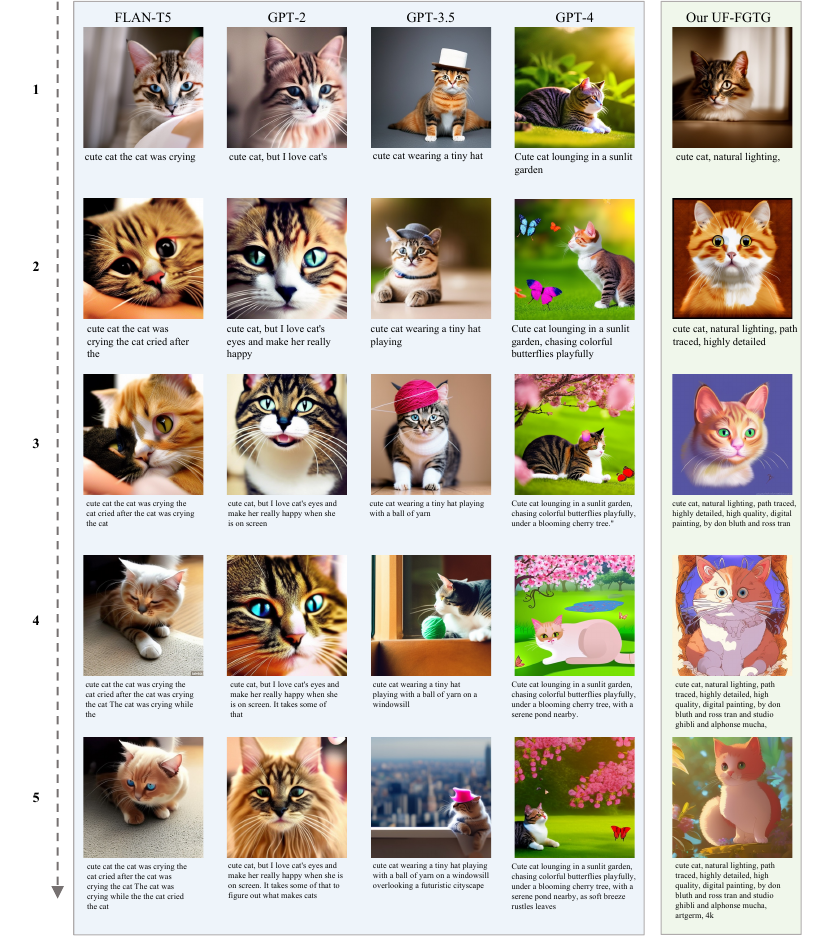}
    }
    \vspace{-1em}
    \caption{Comparison of continuous generation prompts generated by FLAN-T5, GPT-2, GPT3.5, GPT-4, and our UF-FGTG, with corresponding images generated by Stable Diffusion-v2.1 (max token set to 6). Our UF-FGTG generates more reasonable and user-friendly prompts at each step, resulting in visually superior images compared to other methods. Origin prompt: \textit{``cute cat''.}
    }
    \vspace{-1em}
    \label{fig:quantitative_2_6}
\end{figure*}

\begin{figure*}[htb]
    \centering
    \resizebox{\textwidth}{!}{
        \includegraphics[scale=1]{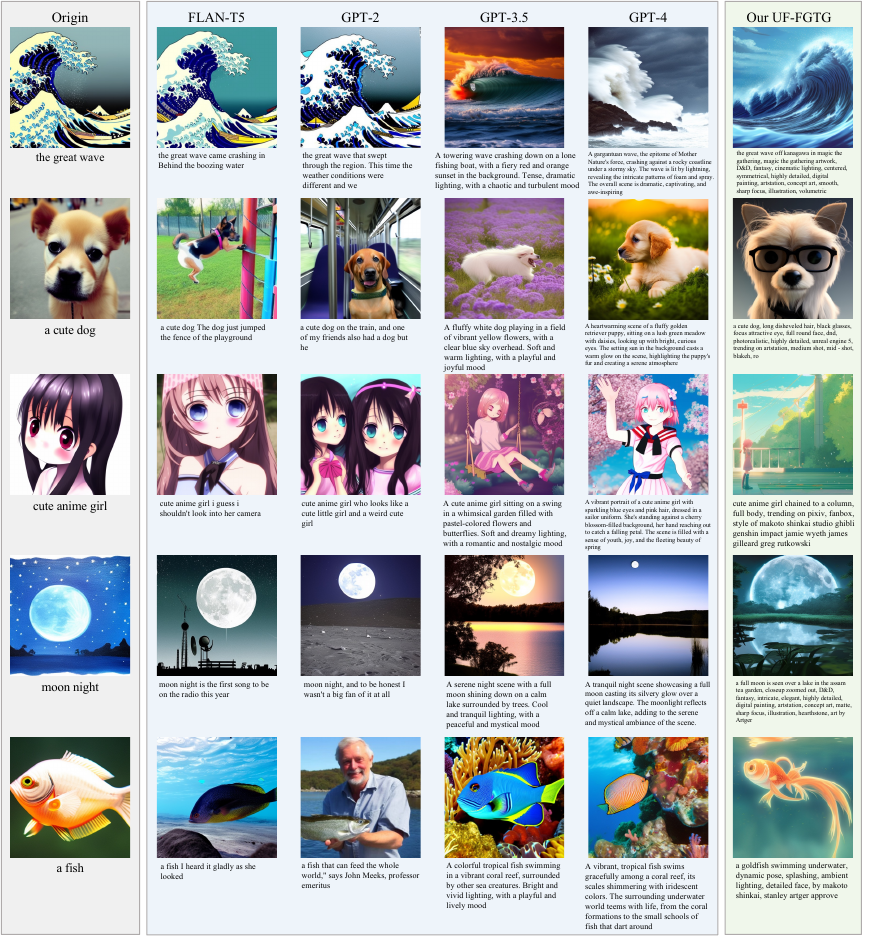}
    }
    \vspace{-1em}
    \caption{ Comparison of prompts generated by FLAN-T5, GPT-2, GPT3.5, GPT-4, and our UF-FGTG, with corresponding images generated by Stable Diffusion-v2.1 (max token set to 20). Our UF-FGTG outperforms other methods significantly in various categories, including animals, humans, landscapes, and plants.
    }
    \vspace{-1em}
    \label{fig:quantitative_2}
\end{figure*}

We present additional visualization results as shown in Fig.~\ref{fig:quantitative_2_6} and Fig.~\ref{fig:quantitative_2}. As mentioned in Sec.~\textbf{Applications}, we propose two strategies for prompt generation. Figure~\ref{fig:quantitative_2_6} illustrates the case of continuous generation with a maximum token setting of 6. Our UF-FGTG is user-friendly and consistently generates better prompts than other methods. Figure~\ref{fig:quantitative_2} presents additional visual results for Sec.~\textbf{Qualitative Comparison}, where the maximum token setting is 20.
\par 
Overall, it can be observed that our UF-FGTG generates prompts that correspond to better images than other methods. Our method achieves superior performance in generating images of various categories, including landscapes, animals, abstract scenes, and human portraits.

\begin{figure}[htp]
    \centering
    \resizebox{\columnwidth}{!}{
        \includegraphics[scale=1]{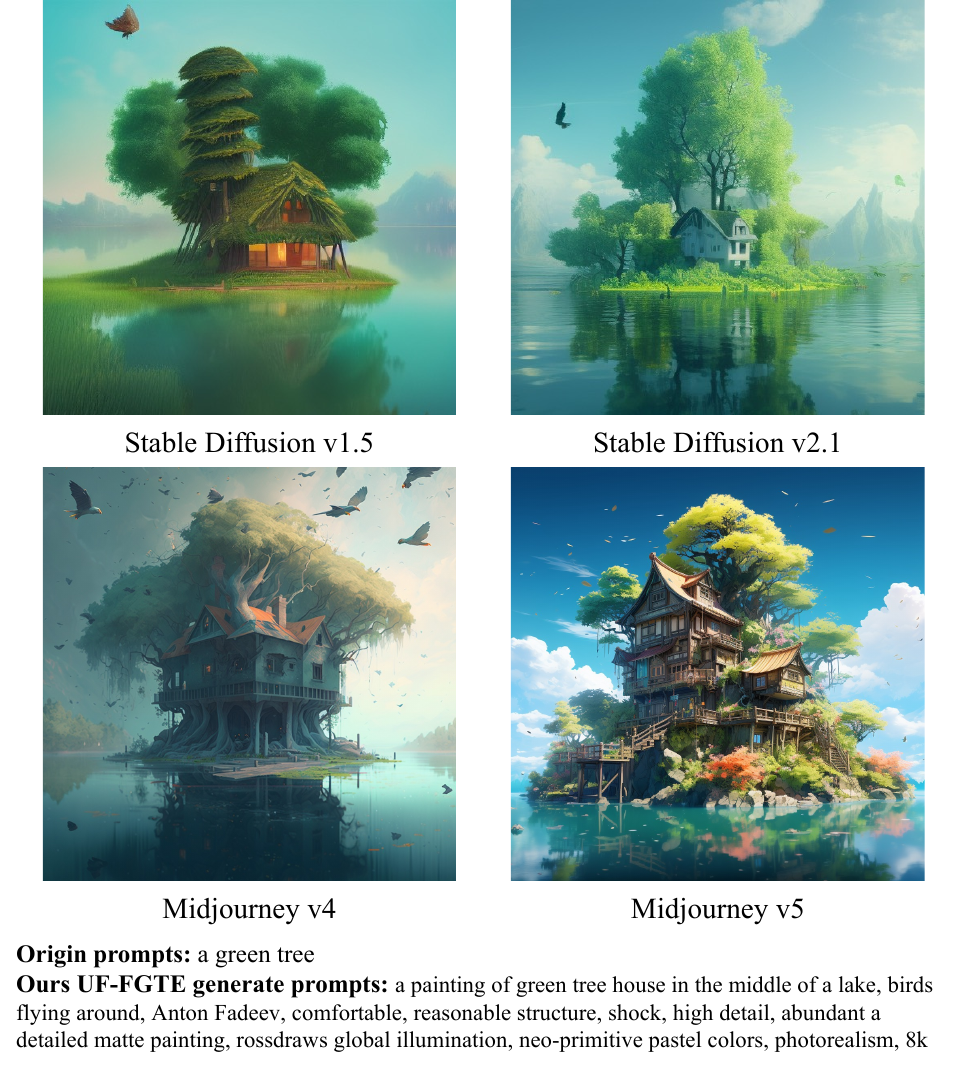}
    }
    \caption{Transferability of UF-FGTG across different text-to-image models.}
    \label{fig:transfer}
\end{figure}
\section{Appendix C: ChatGPT Description}
\label{sec:appendix_c}
The ChatGPT description, also referred to as ``ChatGPT prompt", is a phrase or instruction given to the ChatGPT AI model to generate a response, which can enhance productivity and creativity. 
Following~\cite{chatgpt_diffusion},  we provide the following description for GPT-3.5 and GPT-4:

\par 
\textit{Stable Diffusion is an AI art generation model similar to DALLE-2.}
\par 
\textit{Here are some prompts for generating art with Stable Diffusion. }
\par 
\textit{Example:}
\begin{itemize}
    \item \textit{A ghostly apparition drifting through a haunted mansion's grand ballroom, illuminated by flickering candlelight. Eerie, ethereal, moody lighting. }
    \item \textit{portait of a homer simpson archer shooting arrow at forest monster, front game card, drark, marvel comics, dark, smooth}
    \item \textit{... \textbf{Please see more in~\cite{chatgpt_diffusion}.}}
\end{itemize}
\par 
\textit{The prompt should adhere to and include all of the following rules:}
\begin{itemize}
    \item \textit{Prompt should always be written in English, regardless of the input language. Please provide the prompts in English.}
    \item \textit{Each prompt should consist of a description of the scene followed by modifiers divided by commas.}
    \item \textit{When generating descriptions, focus on portraying the visual elements rather than delving into abstract psychological and emotional aspects. Provide clear and concise details that vividly depict the scene and its composition, capturing the tangible elements that make up the setting.}
    \item \textit{The modifiers should alter the mood, style, lighting, and other aspects of the scene.}
    \item \textit{Multiple modifiers can be used to provide more specific details.}
    \item \textit{After providing the prompt in English, also provide the Chinese translation for each prompt.}
\end{itemize}
\par 
\textit{I want you to write me each prompt exactly about a list of 1 prompt following the rule, and please give me the final 1 prompt.}

\section{Appendix D: Transferability of UF-FGTG}
\label{sec:appendix_d}
Fig.~\ref{fig:transfer} shows our approach's generalizability across various text-to-image models. Our model, initially trained on Stable Diffusion-v2.1, leverages its inherent robustness to generate effective prompts. We hypothesize this robustness allows our model to produce high-quality images even when applied to other models. To verify this, we test our model on Midjourney~\cite{midjourney}, a text-to-image model distinct from Stable Diffusion. Our results confirm that our model generates prompts leading to high-quality images on Midjourney. This suggests our model's prompt-generation ability extends beyond Stable Diffusion to other models.
\par 
Overall, our model's generated prompts can also produce high-quality images on other text-to-image models. This indicates that our method has learned a generalizable representation of text-to-image prompts and can be widely applied to help users generate model-preferred prompts

\end{document}